# Experimental realization of field-induced XY and Ising ground states in a quasi-2D $S$=1/2 Heisenberg antiferromagnet

Yoshimitsu Kohama[1], Marcelo Jaime[1], Oscar E. Ayala-Valenzuela[1], Ross D. McDonald[1], Eun deok Mun[1], Jordan F. Corbey[2], Jamie L. Manson[2]

[1] MPA-CMMS, Los Alamos National Laboratory, Los Alamos, New Mexico 87545, USA

[2] Department of Chemistry and Biochemistry, Eastern Washington University, Cheney, WA 99004, USA

**Abstract**

High field specific heat, $C_p$, and magnetic susceptibility, $\chi$, measurements were performed on the quasi-two dimensional Heisenberg antiferromagnet $[Cu(pyz)_2(pyO)_2](PF_6)_2$. While no $C_p$ anomaly is observed down to 0.5 K in zero magnetic field, the application of field parallel to the crystallographic *ab*-plane induces a lambda-like anomaly in $C_p$, consistent with Ising-type magnetic order. On the other hand, when the field is parallel to the *c*-axis, $C_p$ and $\chi$ show evidence of XY-type antiferromagnetism. We argue that it is a small but finite easy-plane anisotropy in quasi-two dimensional $[Cu(pyz)_2(pyO)_2](PF_6)_2$ that allows the unusual observation of field induced XY and Ising-type magnetic states.

Two dimensional Heisenberg antiferromagnets (2D-HAFM) have been intensely studied on both theoretical and experimental fronts for many years, and continue to be topical due to newly discovered materials. In an early study, Mermin and Wagner [1] demonstrated that strong fluctuations in a strictly 2D model prevent long range ordering at finite temperature. However, the reduction of the spin dimensionality *n* (i.e. the change from Heisenberg (*n*=3) to XY (*n*=2) and Ising models (*n*=1)) suppresses spin fluctuations and leads to different types of transitions and regimes at finite temperatures. If easy-plane anisotropy is introduced, the 2D-HAFM can be described by 2D-XY antiferromagnet (2D-XYAFM) and a Berezinskii-Kosterlitz-Thouless (BKT) transition takes place characterized by a broad peak in the specific heat vs temperature $C_p(T)$.[2,3,4] When easy-axis anisotropy is introduced to 2D-HAFM, the system becomes a 2D-Ising antiferromagnet (2D-IAFM) and shows a second order phase transition characterized by a lambda-type anomaly in $C_p(T)$.[4,5] Since an applied magnetic field can mimic an effective easy-plane anisotropy, as earlier demonstrated, [6-8] the combined effect of external magnetic field and intrinsic easy-axis/easy-plane anisotropy can be used to tune the ground state of HAFM systems. [7-9] However, in most real magnetic systems the inter-plane exchange coupling (*J'*) is generally sufficient to induce 3D ordering, preventing the experimental observation of the crossover from 2D-HAFM to 2D-XYAFM and 2D-IAFM. [10] Hence it is highly desirable to find a system close enough to the 2D limit for the properties and phase transitions to be tuned with external magnetic fields.

In this letter, we provide a remarkable and unprecedented example of both field induced XY and Ising states in a highly anisotropic quasi-2D-HAFM $[Cu(pyz)_2(pyO)_2](PF_6)$. High field specific heat and magnetic susceptibility measurements reveal that the spin anisotropy and resultant nature of the phase transition can be tuned by the orientation of an applied magnetic field relative to the easy-plane. The Hamiltonian describing a 2D-HAFM with finite easy-axis or easy-plane anisotropies in an external field is given by

$$\hat{H} = J/2 \sum_{i,j}[\hat{S}_i^a\hat{S}_j^a + \hat{S}_i^b\hat{S}_j^b + (1-\Delta)\hat{S}_i^c\hat{S}_j^c] - g\mu_B H \sum_i \hat{S}_i^z, \qquad (1)$$

where *J* represents the in-plane antiferromagnetic coupling, Δ is the spin anisotropy, and the sum (*i,j*) is over all nearest neighbors. The isotropic 2D-HAFM corresponds Δ = 0, and the 2D-HAFM with easy-axis and easy-plane anisotropies are Δ < 0 and Δ > 0, respectively. The magnetic field is applied along the *z*-direction, and the last term in Eq. 1 represents the Zeeman energy. If an external magnetic field is applied, the spins align perpendicular to it to minimize the free energy and simultaneously satisfy the AFM exchange interaction, resulting in a strong suppression of spin fluctuations. Thus, when a strong magnetic field is applied along *c*-axis (*z*=*c*), the spin fluctuations along *z* are minimized, and the spin projection (order parameter) in the *ab* plane behaves as XY spin. Although the order parameter can be reduced as field increases, an external magnetic field breaks *O*(3) symmetry in the 2D-HAFM and induces a 2D-XYAFM (as illustrated schematically in Fig.1(a)). Accordingly, Cuccoli et al [6] indicated that the magnetic field mimics an easy-plane anisotropy in pure 2D-HAFM and induces a BKT-like broad $C_p$ peak as field increases (Fig.1(b)).[6] They also predicted that the spin anisotropy, $\Delta$ in eq.1, scales quadratically with the magnetic field as $\Delta \sim 0.1 h^2$, where $h$ is the normalized magnetic field $h \equiv g\mu_B H/(JS)$. In spite of the intense research in this area, [11-17] hitherto the predicted field dependence $\Delta \sim 0.1 h^2$, has never been confirmed, likely due to the non-negligible value of interplane exchange interaction $J'$ in real systems. On the other hand, as theorized many years ago,[8-9] the application of a magnetic field to the easy-plane in the 2D-XYAFM ($\Delta>0$, and $z=a,b$) restricts the rotation to the $ab$-plane and induces an Ising ground state. Since the in-plane spin fluctuations can be tuned by magnetic field strength, the degree of spin fluctuations in two directions ($c$ and $a$, or $b$) becomes similarly weak near the critical field, $H_{Ising}$, at which point the Zeeman energy is equal to the easy-plane anisotropy. In fact, at $H = H_{Ising}$, Eq.1 can be reduced to,

$$\hat{H} = J/2 \sum_{i,j}[\hat{S}_i^b \hat{S}_j^b + (1-\Delta)(\hat{S}_i^a \hat{S}_j^a + \hat{S}_i^c \hat{S}_j^c)]. \tag{2}$$

Equation 2 mimics the easy-axis 2D-HAFM model for $\Delta>0$ (compare with Eq.1)[8], suggesting the emergence of magnetic field induced Ising state in the $H//ab$ case. It is interesting to note that the application of a magnetic field parallel to the easy-axis in 2D-IAFM induces a spin-flop transition [7] which is not anticipated in the 2D-HAFM or XYAFM limits.

[Cu(pyz)$_2$(pyO)$_2$](PF$_6$)$_2$ belongs to a family of isostructual square lattice coordination polymers of general composition [Cu(pyz)$_2$(pyO)$_2$]$X_2$ where pyz is pyrazine (N$_2$C$_4$H$_4$), pyO is pyridine-$N$-oxide (NOC$_5$H$_5$) and $X$ is ClO$_4^-$, BF$_4^-$, or PF$_6^-$ [13-18]. These compounds have either monoclinic ($C2/m$ for $X =$ ClO$_4^-$ and BF$_4^-$) or orthorhombic ($Cmca$ for PF$_6^-$) symmetries. Each Cu$^{2+}$ ion is spin-1/2 and has a tetragonally-elongated stereochemistry. The metal coordination sphere is comprised of four equatorial N-atoms from pyz ligands (Cu-N = 2.045 and 2.067 Å) and two longer axial sites occupied by O-atoms from pyO (Cu-O = 2.317 Å). An extended 2D layer is formed by pyz bridges that link adjacent CuN$_4$O$_2$ octahedra (Cu···Cu = 6.863 and 6.914 Å) into square sheets in the crystallographic $ab$-plane. The $X$ anions required for charge compensation occupy positions between the layers. Excellent 2D magnetic isolation is anticipated owing to the rather large interlayer spacing (closest Cu-Cu = 13.683 Å) provided by the bulk pyO ligands. [Cu(pyz)$_2$(pyO)$_2$](PF$_6$)$_2$ experiences a much smaller $J'$ than $J$ ($J' \sim$ 0.0017 K, $J \sim$8.2 K and $J'/J \sim 2 \times 10^{-4}$) as determined by the experimental observables $H_c^{ab}$, $g^{ab}$ and $T_N$ [18] and the following equations, $g\mu_B H_c=4J+2J'$ [16] and $T_N=0.732\pi J/(2.43-\ln(J'/J))$.[10] The estimated high degree of structural (and exchange) anisotropy suggests [Cu(pyz)$_2$(pyO)$_2$](PF$_6$)$_2$ to be an excellent example of the 2D-HAFM in contrast to other low-dimensional systems such as Cu(pyz)$_2$(ClO$_4$)$_2$ ($J'/J \sim$7 $\times$ 10$^{-4}$ [11]), Cu(tn)Cl$_2$ ($J'/J \sim$ 1 $\times$ 10$^{-3}$ [12]), and [Cu(pyz)$_2$(HF$_2$)]BF$_4$ ($J'/J \sim$ 3 $\times$ 10$^{-3}$ [13]). To date, Sr$_2$CuO$_2$Cl$_2$ is the only quasi-2D HAFM showing an anisotropy larger than the title compound ($J'/J \sim 1\times 10^{-4}$[11]). However, $Sr_2CuO_2Cl_2$ has an in-plane exchange interaction $J$ = 1451 K, which is much too high as compared to current experimental limitations which limits the ability to probe magnetic-field induced phase transitions.

$C_p(H,T)$ and $\chi(H,T)$ were measured on aligned single crystals of $[Cu(pyz)_2(pyO)_2](PF_6)_2$ grown from aqueous solution as described in Ref.18. $C_p$ vs $T$ was obtained using both a standard thermal relaxation technique and a modified relaxation technique known as dual slope method.[19] $C_p$ vs $H$ was measured using an AC technique.[20] $C_p(T,H)$ experiments were carried out in a $^3$He refrigerator furbished with a 15 T superconducting magnet, and in a 50 T pulsed magnet equipped with a $^4$He cryostat. The $\chi(H,T)$ experiments were performed with a Physical Property Measurement System® manufactured by Quantum Design, Inc. The magnetic contribution to the specific heat, $C_m(T)$ was obtained by subtracting the lattice specific heat estimated from high temperature data and analyzed in a similar fashion as in Ref.17.

Figure 1(c) and (d) show $C_m(T)$ for several magnetic fields applied parallel and perpendicular to the 2D magnetic planes. In the absence of an applied field we observe a smooth, featureless magnetic contribution, as expected for highly 2D systems. Indeed, the Monte Carlo simulations (black curve in Fig.1(b)) [6,21] indicate no features in pure 2D-HAFM. Accordingly, we find that $C_m$ follows the predicted power-law behavior for pure 2D-HAFM, $C_m \sim aT^2+bT^4$ in the low temperature limit (inset of Fig.1(d)).[22]

The application of magnetic field induces features in $C_m$, and these features change shape according to the field intensity and orientation. For $H//c$ (i.e. normal to the 2D magnetic planes) broad peaks were observed. These broad peaks become much more prominent with increasing magnetic field, while the peak temperature first increases and then drops above ~7 T. The shape and increasing intensity of the peak in the high field region agree with the previous Monte Carlo results in pure 2D-HAFMs (Fig.1(b))[6], indicating that the system show a field-induced XY behavior in this field orientation. It is important to note that the less anisotropic 2D-HAFMs, $[Cu(pyz)_2(HF_2)]BF_4$ ($J'/J \sim 3 \times 10^{-3}$)[13] and $[Cu(pyz)_2(HF_2)]PF_4$ ($J'/J \sim 6.3 \times 10^{-1}$)[14] show sharp $C_m$ peaks, which is evidence of 3D-ordering temperature (Néel transition), on top of the BKT-like broad peak in all relevant magnetic fields. The absence of the sharp peak in the title compound likely a direct consequence of the extremely high anisotropy.

When a weak magnetic field (<5 T) is applied in the *ab* plane, a characteristic λ-like peak is observed (Fig.1(c)). This λ-like anomaly can be seen on the low-temperature side of the broad peak. The λ-like peak for $H//ab$ is clearly observed in high resolution/high sensitivity $C_m(H)$ measurements performed using an AC technique (Fig.1(e) and Fig.1(f) for $H//ab$ and $H//c$, respectively). Indeed, at low fields the difference in $C_m(H)$ between $H//ab$ and $H//c$ is remarkable, where the $C_m$ for $H//c$ show only a shoulder-shaped anomaly. The field-induced λ-shaped anomaly observed for $H//ab$ is characteristic of Ising-type ordering, and seems to evolve into a BKT-type broad anomaly at higher fields. Although 3D-ordering can lead to a similar sharp peak, it is not expected in a very high anisotropy sample as $Sr_2CuO_2Cl_2$.[21] Additionally, a 3D Néel transition does not show field orientation dependence, whereas the absence of a sharp peak for *H//c* is inconsistent with such a transition. In addition, a field-induced Ising state can be expected from Eq.2 when an external magnetic field is applied to *ab*-plane.[8] Thus, we conclude that the sharp peak is the signature of Ising-like transition. Here, we would like to note that the Monte Carlo method cannot be carried out for magnetic field applied for *H//ab* due to the well-known sign problem. Consequently, further theoretical development is necessary for reproducing the experimental data at a quantitative level. At *H*~25 T we also see a broad anomaly in $C_m(H//ab)$. This broad anomaly arises from thermal excitations between magnetic spin levels, *i.e. a* Schottky anomaly, corresponding to the magnetization *M(H)* saturation at ~24T.[18]

The magnetic contribution to the specific heat is determined by calculating the difference $\Delta C_p(T,H) = C_p(T,H) - C_p(T,0)$ and we plot $\Delta C_p T^{-1}$ in Fig.2. Previous Monte Carlo simulations have shown that the magnitude of $\Delta C_p T^{-1}$ monotonically increases with magnetic field.[6,23] A clear confirmation of the prediction is seen in Fig.2(a). To do a quantitative comparison, we need to take into account the easy-plane anisotropy which is $\Delta$~0.007 as evaluated later. From the expression $\Delta$~$0.1 h_p^2$ [6] we introduce the effective magnetic field $h_{eff} = h + h_p$, where $\Delta$ plays the role of an internal magnetic field $h_p$~0.26. The inset of Fig.2(a) shows the magnitude $\Delta C_p T^{-1}$ as a function of $h_{eff}$. We find a good agreement with the theoretical prediction below $h_{eff}$~2 (*H*~3.9T), while the higher field data separates from the theory. The observed departure is attributed primarily to the change in spin-band structure as a function of field.[24] In other words, the external magnetic field alters the background contribution to $C_p$ which is due to spin-band structure. In any case, below *h*~2, the numerical results agree well with our data, providing strong evidence for field-induced XY antiferromagnetism. In contrast, Fig.2(b) reveals a λ-like peak when a weak magnetic field (*H* ≤ 3T) is applied in the *ab*-plane. Below 2T, the peak height is roughly twice as high as the BKT broad peak observed for *H//c*. However, above 5T, the peak height is almost identical to the *H//c* data. Since the order parameter (the XY component of spin) is reduced as field increases, this points that the λ-like peak can be weaken with decreasing the order parameter.

Figure 3 shows the DC susceptibility $\chi(T) = M/H$, for *H//c* and *H//ab*. In the low field region, $\chi$ shows a broad bump around 7.7 K which is characteristic of the 2D-HAFM.[22] The rounded maximum at a temperature, $T_{max}$, relates to the onset of the antiferromagnetic short range ordering, resulting in the reduction of $\chi$ at lower temperature. Various theoretical studies indicate that $T_{max}$ is given approximately by the in-plane exchange constant *J*.[4,6,22] This estimate roughly coincides with the independent estimation of $J \approx 8.2$ K from $H_c$, $T_c$, and *g*-factors. For both field orientations, an upturn is observed below *T*=3 K. According to Monte Carlo simulations,[6] the minimum temperature ($T_{min}$) in $\chi(T)$ marks the onset of XY behavior below $2h_{eff}$ (~3.9T). A similar minimum is also observed for $H//ab$, and both $T_{min}$ for $H//ab$ and $H//c$ occur at temperatures slightly higher that the anomaly in $\Delta C_p$ as indicated by arrows in Fig.3. This behavior is expected for 2D-XYAFM and 2D-IAFM in the low field region.[4,6] The derivative $\partial\chi(T)/\partial T$ is plotted for both field orientations in the insets of Fig.3(a) and 3(b). While a sharp peak is seen for $H//ab$, just a broad feature is evident for $H//c$. We interpret the sharp peak as arising from the Ising nature of the magnetic transition. As the magnetic field is increased, the peak becomes smaller and the difference between field orientations vanishes. Indeed, $\partial\chi(T)/\partial T$ at $H = 5T$ is similar for both $H//c$ and $H//ab$. As in the case of $C_m(T,H)$, a strong enough magnetic field reduces the amplitude of the order parameter and makes difficult to observe the field orientation difference between $H//c$ and $H//ab$.

A quasi-2D system with low anisotropy shows anomalies in both $C_p$ and $\chi$.[14] By contrast, our data reveal a disappearance of the $\chi(T)$ anomaly in the high field region in marked contrast to the large BKT peak observed in $C_m(T)$, which grows with $H$. This behavior was identified in earlier Monte Carlo simulations as a signature of the magnetic field-induced 2D-XYAFM,[6] which can be understood from a microscopic point of view. While the peak in $C_m(T)$ relates to the magnetic entropy, *i.e.* it is a measurement of the magnetic degrees of freedom in all directions, $\chi(T)$ measures the fraction of spins that are tilted in the applied field direction. In principle, the spin $z$-component cannot fluctuate in high fields and the vortex/antivortex creation in the BKT transition is the ordering perpendicular to $z$ axis which cannot induce any anomaly in $\chi$ ($z$ spin component), but can change the degrees of freedom in the XY plane. This explains why our data show an obvious $C_m(T)$ anomaly and no $\chi(T)$ anomaly in the high field region. Hence, our complementary measurements of $C_p$ and $\chi$ strongly support the magnetic field induced 2D-XYAFM state in $[Cu(pyz)_2(pyO)_2](PF_6)_2$.

Figure 4 displays the $T$-$H$ phase diagram as obtained from our $C_p(T,H)$ measurements. A clear non-monotonic dependence of $T_p$ with respect to field was found which is similar to other quasi-2D systems.[13,14] Sengupta *et al.*[13] proposed that the non-monotonic behavior is caused by the phase fluctuations typical for a 2D system. The inset in Fig.4 compares the experimental $T^{exp}_{min}$ collected for $H//c$ to the theoretical $T^{theory}_{min}$.[6] Here, the experimental $T^{exp}_{min}$ is plotted as a function of $h_{eff}$. If we assume no easy-plane anisotropy ($\Delta$, $h_p =0$), $T^{exp}_{min}$ shows a clear departure from the theory. However, if we take $\Delta=0.007$ ($h_p=0.26$), the agreement between $T^{exp}_{min}$ and $T^{theory}_{min}$ becomes significantly better. This value of the spin anisotropy is in good agreement with independent microwave frequency measurement of antiferromagnetic resonance in this compound.[25] Mean field analysis of the antiferromagnetic resonance estimates $\Delta=0.003$, which is an under estimate compared to finite cluster analysis.[14,15] The observed agreement confirms that the easy-plane anisotropy acts as an external magnetic field, and *vice versa*.

In summary, we have studied the field-induced XY and Ising ground states in the $S = 1/2$ weakly easy-plane quasi-2D HAFM [Cu(pyz)$_2$(pyO)$_2$](PF$_6$)$_2$ with $C_p(T,H)$ and $\chi(T,H)$ measurements. Since the external magnetic field mimics an additional easy-plane anisotropy, for $H//$c, the system then displays an XY ground state. On the other hand, when the magnetic field is applied parallel to the *ab*-plane, by the combination of the intrinsic easy-plane anisotropy and the external magnetic field, an Ising ground state emerges. Finally, we emphasize that the field-induced behavior reported here very likely arises from the extreme two dimensionality: an extremely weak $J'$ compared to $\Delta$ ($J'/J \sim 2\times10^{-4} \ll \Delta \sim 0.007$) in [Cu(pyz)$_2$(pyO)$_2$](PF$_6$)$_2$. In contrast, the less anisotropic system [Cu(pyz)$_2$(HF$_2$)]PF$_6$ ($J'/J \sim 0.03$ and $\Delta \sim 0.003$ [14]) shows a sharp anomaly in $C_p(T)$ for all values and orientation of external magnetic field, signature of the traditional Néel transition. The first observation of field-induced 2D-XYAFM and 2D-IAFM physics is now unambiguously demonstrated in [Cu(pyz)$_2$(pyO)$_2$](PF$_6$)$_2$.

We acknowledge fruitful discussions with T. Roscilde, C.D. Batista, and J. Singleton. Y.K., M.J., E.M., O.A. and R.M. are supported by the National Science Foundation, the Department of Energy, and the State of Florida. Work at EWU was supported by the National Science Foundation under Grant No. DMR-1005825.

25. O.E. Ayala-Valenzuela *et al*., in preparation.

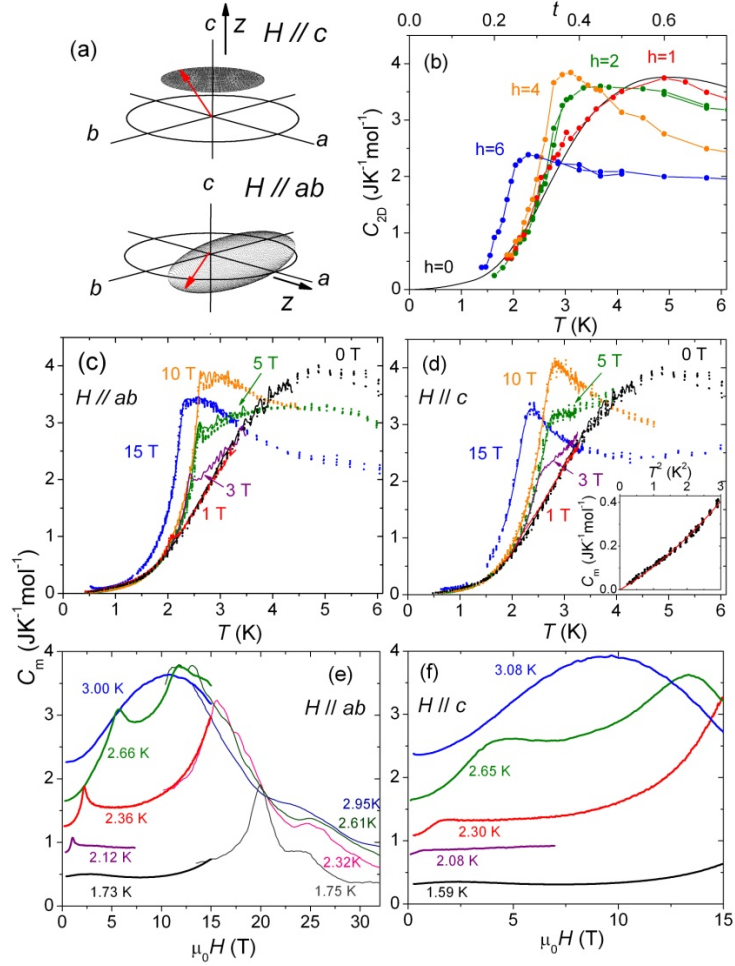

Fig. 1. (color online) (a) Schematic illustrating the spin configuration of the field-induced XY (top), and field-induced Ising (bottom) states in weakly easy-plane HAFM. The black circles and red arrows are the easy-plane and spin respectively. The shadowed surface represents the direction at which spin can point out. In the case of isotropic HAFM, it consists of spherical shape. With finite Δ, the surface forms pancake-shape. When strong magnetic field is applied perpendicular to the easy-plane of a 2D-HAFM, the surface becomes disk-like shape and projection of spin on the *ab*-plane behaves as XY spin (top). If weak magnetic field is in the easy-plane, the surface becomes cigar-shaped, and the system can be approximated by the 2D-IAFM (bottom, see the text). (b) Predicted magnetic specific heat $C_{2D}(T)$ for a pure 2D-HAFM when *H//c*. The curves [6,23] are calculated by means of a quantum Monte Carlo algorithm. The normalized magnetic fields of *h*=1,2,4,and 6 corresponds to the magnetic field of 2.0, 4.7, 10.1, and 15.5 T for *H//c*. (c,d) Experimental specific heat $C_m(T)$ for *H//ab* and *H//c*. Here, data collected using the dual slope method are plotted by solid curves, while the data measured by relaxation method are plotted by dots. The inset of Fig.1(d) shows $C_m$ vs $T^2$ below 1.7 K. (e,f) $C_m(H)$ for *H//ab* and *H//c*. The field sweep data at almost constant temperature were measured by means of AC technique in DC and pulsed magnets. These data were normalized to the data measured by relaxation method. The uncertainties of temperature during field sweep are ±0.02K for DC field and ±0.05K for pulsed field experiments.

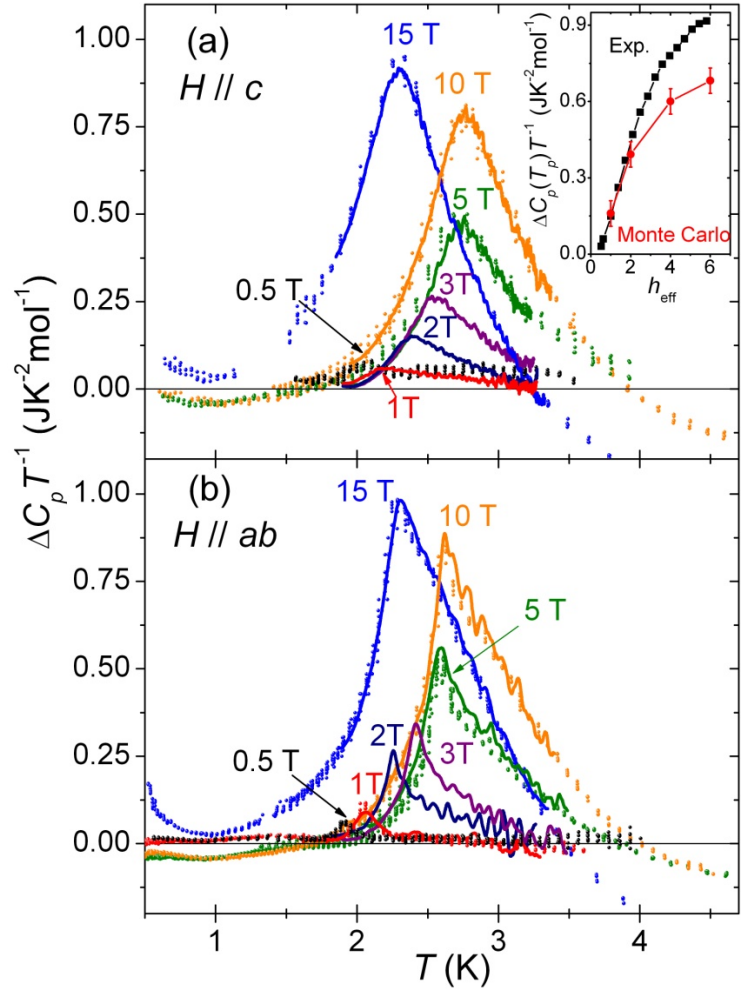

Fig.2. (Color online) Difference of specific heat between finite and zero magnetic field. The curves and dots are the data obtained with Dual slope method and traditional relaxation technique, respectively. The upturn of $\Delta C_p(15T)T^{-1}$ below 1K likely comes from a magnetic nuclear Schottky contribution to $C_p$. The inset shows the peak height of $\Delta C_p/T$ as a function of effective field.

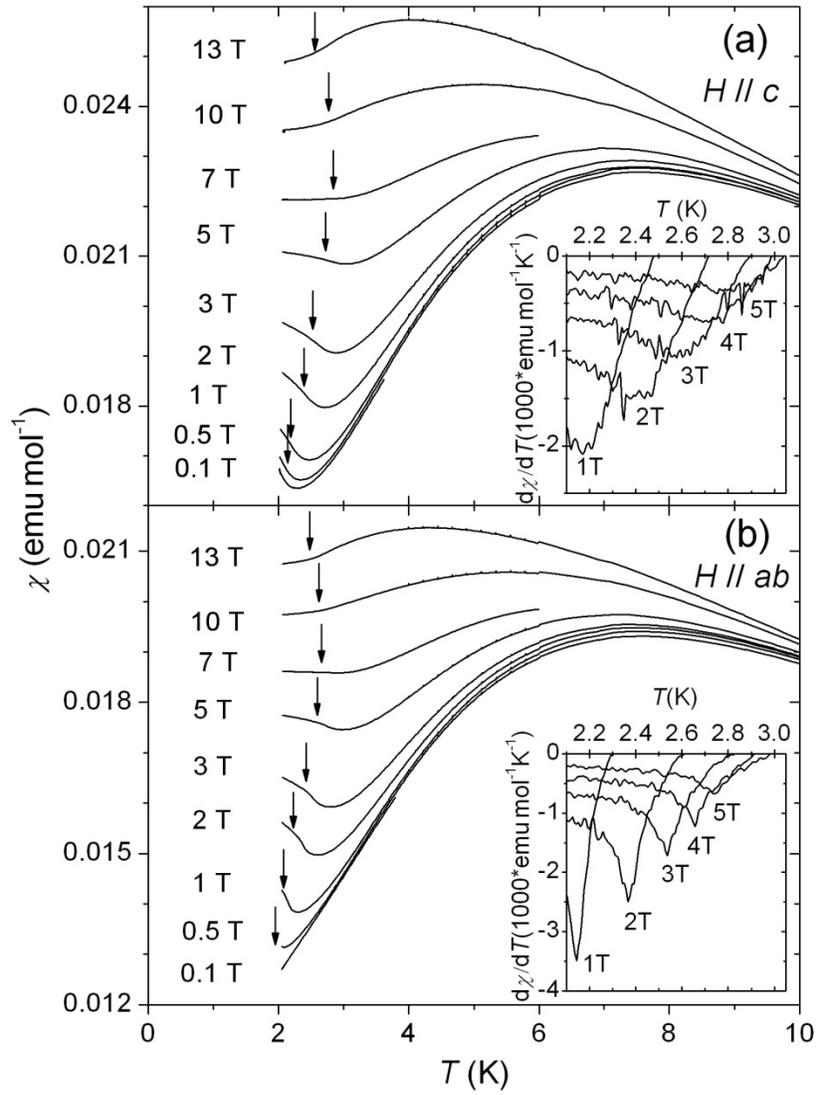

Fig. 3. Temperature dependence of $\chi$ at selected magnetic fields. The arrows indicate the peak temperature in $\Delta C$. The different magnitude of $\chi$ observed for $H//c$ and $H//ab$ is due to the anisotropic $g$-factor.[14] The insets show the derivative of $\chi$ below 5T.

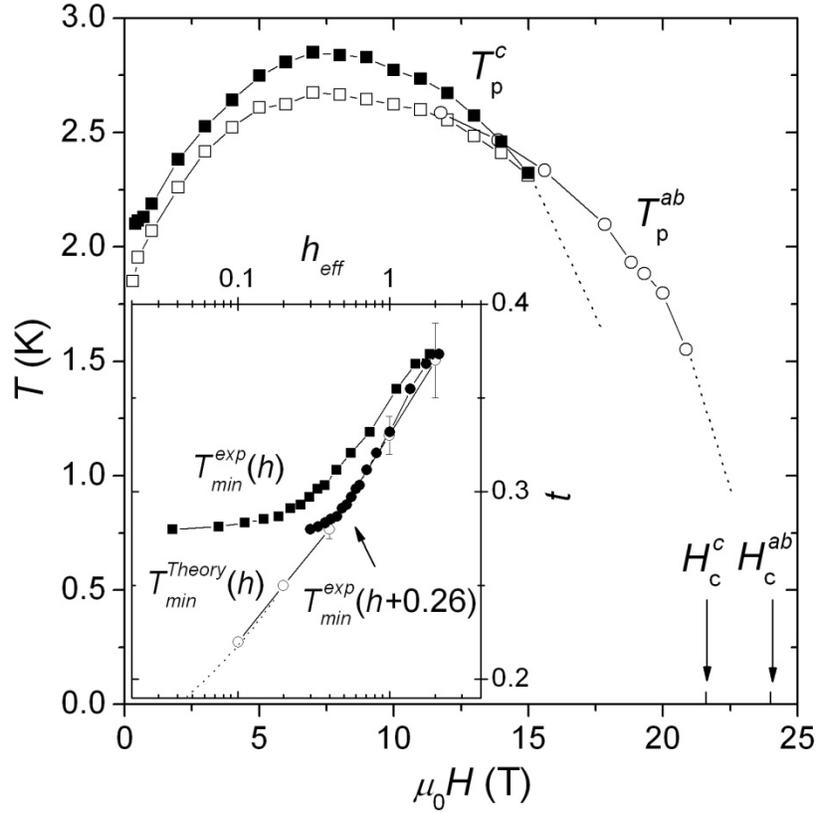

Fig. 4. Magnetic field vs Temperature phase diagram for $H//c$ and $H//ab$ orientations. The solid and open symbols are the $T_p$ for $H//c$ and $H//ab$. The open circles and squares are $T_p$ determined by $C_p$ measurements in pulsed and DC magnets, respectively. $H_c$ values are independently estimated by magnetization measurement [18]. The inset compares our experimental $T_{min}$ to theory. The horizontal and vertical axis are the effective magnetic field (see text) and normalized temperature ($t \equiv J/K= 8.2K$). The open circle is the $T_{min}$ from Monte Carlo simulation.[6,23] The solid circles and squares are the experimental $T_{min}(h_{eff})$ with/without taking into account easy-plane anisotropy.